\documentclass[a4paper,10pt,aps,pre,twocolumn,superscriptaddress,showpacs,
amsmath,amssymb,nofootinbib]{revtex4}
\usepackage{graphicx,color}

\newcommand{\dd}{\mathrm{d}}

\newcommand{\IInt}[3]{\int_{#2}^{#3}\dd #1\;}

\renewcommand{\vec}[1]{\mathbf #1}

\newcommand{\vhi}{\varphi}

\newcommand{\x}{\vec r}
\newcommand{\Dr}{D_\text{r}}
\newcommand{\Te}{T_\text{eff}}
\newcommand{\vc}{v_\text{c}}

\newcommand{\nois}{\boldsymbol\xi}

\begin{document}

\title{Active colloidal suspensions: Clustering and phase behavior}
\author{Julian Bialk\'e}
\email{jubia@thphy.uni-duesseldorf.de}
\affiliation{Institut f\"ur Theoretische Physik II,
  Heinrich-Heine-Universit\"at, D-40225 D\"usseldorf, Germany}
\author{Thomas Speck}
\affiliation{Institut f\"ur Physik, Johannes Gutenberg-Universit\"at Mainz,
Staudingerweg 7-9, D-55128 Mainz, Germany}
\author{Hartmut L\"owen}
\affiliation{Institut f\"ur Theoretische Physik II,
  Heinrich-Heine-Universit\"at, D-40225 D\"usseldorf, Germany}
\begin{abstract}
  We review recent experimental, numerical, and analytical results on active
  suspensions of self-propelled colloidal beads moving in (quasi) two
  dimensions. Active colloids form part of the larger theme of \emph{active
    matter}, which is noted for the emergence of collective dynamic phenomena
  away from thermal equilibrium. Both in experiments and computer simulations,
  a separation into dense aggregates, i.e., clusters, and a dilute gas phase
  has been reported even when attractive interactions and an alignment
  mechanism are absent. Here, we describe three experimental setups, discuss
  the different propelling mechanisms, and summarize the evidence for phase
  separation. We then compare experimental observations with numerical studies
  based on a minimal model of colloidal swimmers. Finally, we review a
  mean-field approach derived from first principles, which provides a
  theoretical framework for the density instability causing the phase
  separation in active colloids.
\end{abstract}


\maketitle


\section{Introduction}

In the past decade, active systems have gained enormous interest in the field
of soft matter physics from both the experimental and the theoretical side,
see Refs.~\citenum{vicsek12,marc13,Aranson13,marchetti_arxiv} for recent general
reviews. Motivated through macroscopic biological systems like flock of
birds~\cite{cavagna10} and school of fish~\cite{becco06}, but also microscopic
systems like bacterial colonies \cite{peruani12,zhang10}, theoretical models
of self-propelled particles have been developed that demonstrate the emergence
of collective phenomena from simple idealized
interactions~\cite{vicsek95,romanczuk08}. Theoretical descriptions have mainly
focused on hydrodynamic approaches describing the coarse-grained dynamics on
large scales~\cite{toner98}. Coefficients are either treated as free
parameters or are derived, e.g., from the microscopic modeling of
collisions~\cite{bert06,ihle12,frey13}. In these models, the crucial
interaction mechanism responsible for collective behavior such as laning,
swarming, and even active turbulence~\cite{wensink12,wensink12_jpcm,farrell12}
is the alignment of velocities, or orientations. These interactions might be
cognitive as in the case of birds, or physical due to, e.g., volume exclusion
of granular rods~\cite{nara07} and disks~\cite{dauchot10}.

\begin{figure*}[t]
  \centering
  \includegraphics{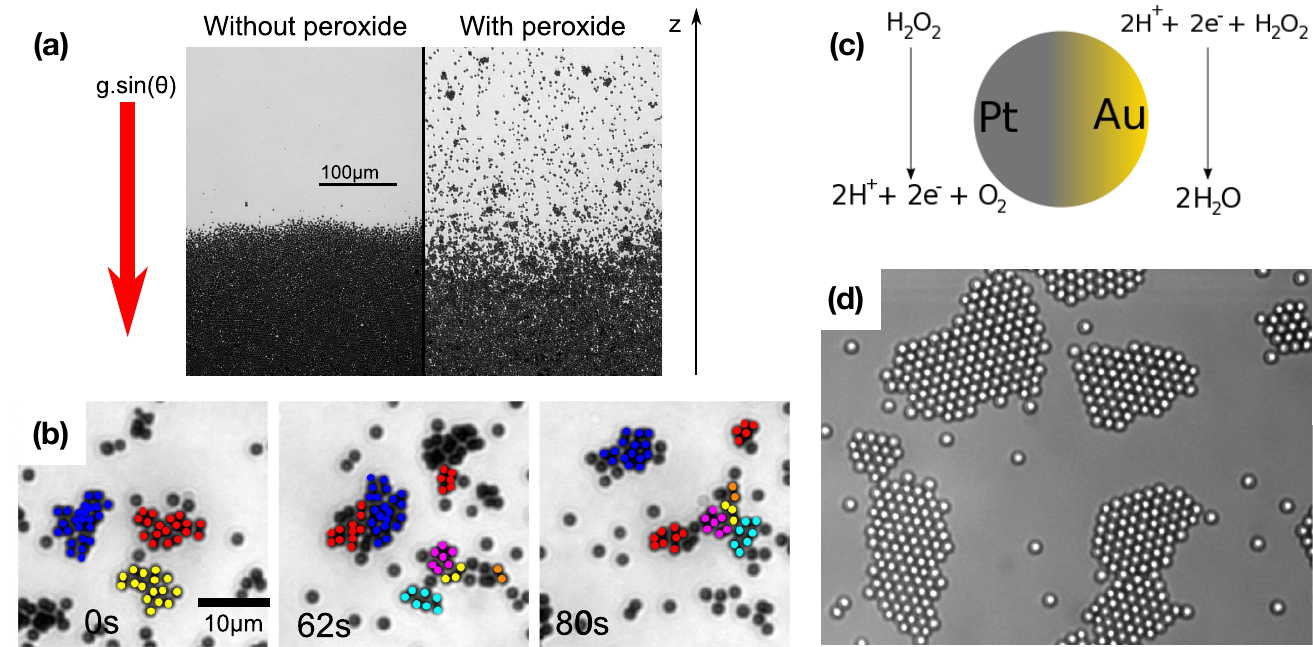}
  \caption{Suspensions of catalytic Janus particles close to a surface:
    (a)~Snapshots of platinum coated gold particles without (left) and with
    ``fuel'' (right) in the experiments of Theurkauff~\emph{et
      al.}~\cite{theur12}. Particles have sunk to the bottom of a tilted cell,
    where they accumulate at the bottom. In the active suspension (right), a
    smeared interface between a dense phase at the bottom and a dilute gas
    phase at the top is observed. (b)~Cluster formation in the dilute
    phase in the experiment of Theurkauff~\emph{et
      al.} Colors indicate membership of a cluster at
$t=0$ and demonstrate
    how clusters evolve. (c)~The platinum acts as a catalyst for the
    decomposition of hydrogen peroxide. The actual swimming mechanism is still
    somewhat debated, see text. (d)~Formation of large clusters, ``living
    crystals'', in a related experiment performed by Palacci~\emph{et
      al.}~\cite{pala13} using colloidal particles with an embedded hematite
    cube. The catalytic activity of the hematite is controlled externally
    through light. Figure adapted from Refs.~\citenum{theur12,pala13}.}
  \label{fig:exp}
\end{figure*}

More recently, experimental setups of artificial colloidal ``swimmers'' have
been realized, the propulsion properties of which can be tuned. Directed
phoretic motion of these colloidal particles is the hydrodynamic consequence
of maintaining a local gradient of a molecular solvent, e.g. due to chemical
reactions on the different surface areas of a particle in a hydrogen peroxide
mixture~\cite{paxton06,hong07,pala10}, or the local demixing of a water-lutidine
mixture at one side of the particle~\cite{volpe11}. Moderately dense
\emph{active suspensions} of such
artificial swimmers can be realized and studied~\cite{theur12,pala13}, for a
summary of the experiments see Fig.~\ref{fig:exp}. Arguably the most
interesting feature is that a clustering of particles is observed. These
clusters are very dynamic, and particles join and leave as shown in
Fig.~\ref{fig:exp}(b). While the cluster size in these experiments seems to
reach saturation, in another experiment~\cite{butt13} using the reversible
demixing of a binary solvent evidence for phase separation into compact large
clusters and a dilute gas phase of free swimmers has been presented.

Such a phase separation has also been observed in computer simulations of a
minimal
model~\cite{fily12,sten13,redner13_prl,butt13,bialke13,sten14,fily14,speck14}.
In
this model, disks are propelled with constant velocity along their
orientations, which undergo free rotational diffusion. Moreover, disks
interact via a purely repulsive pair potential. The existence of a collective
phase transition is somewhat surprising given that this model lacks both
attractions -- leading to phase separation in passive suspensions -- and an
alignment mechanism. Still, the persistence of the directed motion in
combination with volume exclusion forces leads to a self-trapping phenomenon,
where particles get temporally ``stuck'' and block each other, which also has
been shown for lattice models before~\cite{thomp11,soto14_pre}.
Tailleur and
Cates have shown theoretically for a model of run-and-tumble bacteria that
indeed a locally reduced mobility is sufficient to give rise to a separation
into dense \emph{slow} regions, where directed motion is blocked, and a dilute
gas of fast particles~\cite{tailleur08,cates10,cates13}.

Instead of giving a general overview, in this article we focus on recent
experimental and theoretical progress on the phase behavior of self-propelled
colloidal particles in two dimensions without an alignment mechanism. First,
we review results from three groundbreaking experimental setups that have
realized (quasi-)two-dimensional systems of spherical swimmers with
controllable propelling speed $v_0$ of the order $\mu$m/s, where the
correlation between the particle orientations, i.e., the direction of
propulsion, appears to be negligible, see Supplementary Material of
Ref.~\citenum{butt13}. A minimal model is then described, which
nevertheless captures the relevant ingredients of the experiments. We discuss
numerical results based on this model and compare them to experimental
results. We briefly discuss the influence of hydrodynamic interactions as well
as freezing of active systems at high densities. Finally, we introduce a
mean-field approach leading to evolution equations for the density and the
orientational field of an active suspension~\cite{bialke13}. The crucial role
in this theory is played by a single parameter, the force imbalance due to an
anisotropic pair distribution.
We then conclude and outline possible directions for further research in this
rapidly evolving field.


\section{Experimental evidence}

\subsection{Clustering of catalytic swimmers}

For colloidal particles to ``swim'' autonomously, at least the following two
conditions need to be met: (i)~besides the colloidal solute and the solvent,
there is a molecular solute and (ii)~the distribution of this molecular solute
is kept asymmetric\footnote{Thermophoresis
  could in principle also work~\cite{jian10,gole12}, but the required high
  illumination powers induce optical forces, which, in the context considered
  here, are less desirable.}. Two practical schemes have been realized for the
study of (moderately) dense active suspensions: the decomposition of water
peroxide~\cite{paxt04} and the reversible, spinodal demixing of a binary
water-lutidine solvent~\cite{butt12}.

While aggregation of catalytic swimmers has been observed
before~\cite{hong07}, clusters of active colloids have been characterized the
first time in experiments performed by Theurkauff~\textit{et
  al.}~\cite{theur12}. They prepared so-called Janus particles consisting of
two surfaces with different physical properties. In this particular experiment
they used spherical gold particles with one hemisphere coated with
platinum. Immersing these particles in a solvent containing hydrogen peroxide
H$_2$O$_2$, the particles are propelled along their symmetry axis. The
propulsion is realized due to the different chemical properties of platinum
and gold, leading to different rates of H$_2$O$_2$ consumption, see
Fig.~\ref{fig:exp}(c). The mechanism actually responsible for the propulsion
(diffusiophoresis, electrophoresis, or a combination of both) is still
somewhat debated, see Ref.~\citenum{brown14} for a more detailed account for
polysterene-Pt swimmers.  At sufficient low concentrations of hydrogen
peroxide, the propelling speed is proportional to the H$_2$O$_2$
concentration. Of course, at some point, the swimming velocity saturates due
to the finite number of active sites on the particle
surface~\cite{howse07}. The swimming motion of a single particle, as measured
by the mean-squared displacement, fits excellently with the prediction of a
simple theoretical model~\cite{pala10,ebbens10,ke10}, which is discussed in
Sec.~\ref{sec:model}. The experiment can even be performed at high densities
since particles do self-propel at H$_2$O$_2$ concentration below $0.1\%$,
which, in addition, prevents the creation of unfavorable O$_2$ bubbles.

In order to realize different density regimes, Theurkauff~\textit{et al.} have
confined particles in a slightly tilted cell, which creates a reduced gravity
field. The resulting sedimentation profile is more stretched compared to the
equilibrium case, giving the possibility to study the system at different
densities corresponding to different heights in one single sample, see
Fig.~\ref{fig:exp}(a). At low to intermediate densities, the suspension shows
the formation of several clusters, cf. Fig.~\ref{fig:exp}(b). Once clusters
are formed, particles do not stay in their initial cluster but are
continuously exchanged between clusters, see Fig.~\ref{fig:exp}(b). For a
better understanding of this cluster phase, the structure factor
has been measured, which shows that clusters are highly ordered with pronounced
peaks at values of the wave vector $k$ corresponding to the hexagonal lattice.
Simultaneously, an apparently diverging behavior for $k\rightarrow 0$ is
observed, which has been the first experimental indication of density
fluctuations at large length scales for self-propelled beads. This observation
is typical for systems exhibiting finite cluster phases, which can also be seen
for passive colloids with attractive interactions~\cite{cates04}.

Further studies at intermediate densities for packing fractions
$\phi=0.03-0.5$ in a nontilted cell show a linear correlation between mean
cluster size and the average velocity of the particles. This is corroborated
by a theoretical description based on the chemotactic Keller-Segel
model~\cite{keller70}. The use of this model is justified through the fact
that each particle creates a monopole field of H$_2$O$_2$ or O$_2$ around
itself, which acts as chemoattractant for nearby particles. One of the solutions
of the model includes a collapse of the structure into dilute and dense regions
\cite{brenner98}. Although the model does not provide a description for the
kinetics of the clusters, the threshold for this collapse, i.e., the number of
particles in a dense region, is shown to be proportional to the particle
velocity in agreement with the mean cluster size in the experiments.

In the second experiment, Palacci~\textit{et al.}~\cite{pala13} have performed
experiments on catalytic colloidal swimmers, in which the propulsion can be
controlled by light. The particles consist of an antiferromagnetic hematite
cube enclosed by a polymer sphere in such a way that a part of the hematite
cube is exposed to the solvent. When particles are again immersed in a solvent
mixture containing H$_2$O$_2$ the system is in thermal equilibrium in case of
bright-field illumination. As soon as the suspension is illuminated by
blue-violet light (430 to 490 nm), particles can be described as
two-dimensional swimmers. The mechanism behind the propulsion is that the
blue-violet light triggers the chemical decomposition of hydrogen peroxide at
the exposed part of the hematite cube. In addition the hematite cube instantly
points downwards and propels the particles toward the cells bottom. The
colloids then surf on the induced osmotic flow and their motion is captured by
the model of self-propelled Brownian particles in two dimensions, which will be
introduced in Sec.~\ref{sec:model}. The
experiment shows the formation of a few big crystalline clusters just like the
interchange of particles between different clusters, see
Fig.~\ref{fig:exp}(d).

Furthermore, Palacci~\textit{et al.} show that the system exhibits a transition
regarding the number fluctuations $\Delta N\propto N^\alpha$, where the exponent
changes from its equilibrium value $\alpha=1/2$ to giant number fluctuations
with exponent $\alpha\approx0.9$ at $\phi\approx0.07$. Note that these giant
number fluctuations are nothing specific for active systems, since $\alpha=1$ is
expected for any phase separating system. One central result of this
experiment is the reversibility of the cluster phase. Once clusters are formed
and the illumination is turned off subsequently, one observes that all clusters
dissolve, which shows the absence of equilibrium attractions that are
sufficiently strong to induce accumulation of particles. However, the authors
report a strong phoretic attractive force when particles are active. This is
demonstrated by analyzing the radial velocity $v_r$ between particle pairs
showing the relation $v_r\sim r^{-2}$ which is characteristic for phoretic
attraction. In order to show that the activity and not the phoretic attraction
is responsible for the clustering, one can apply an external magnetic field,
causing all particles to propel themselves in the direction of the magnetic
field. It is shown that such a directed non-diffusive propulsion is not
sufficient to maintain a given cluster, because particles drift apart due to
diffusion, i.e., phoretic attraction is not strong enough. As soon as the
magnetic field is turned off and illumination is turned on again, the cluster
reforms. This demonstrates experimentally that clustering of self-propelled
beads is caused by a self-trapping mechanism that essentially depends on the
combination of both self-propulsion and rotational noise.

\subsection{Phase separation}

\begin{figure}[b]
  \includegraphics{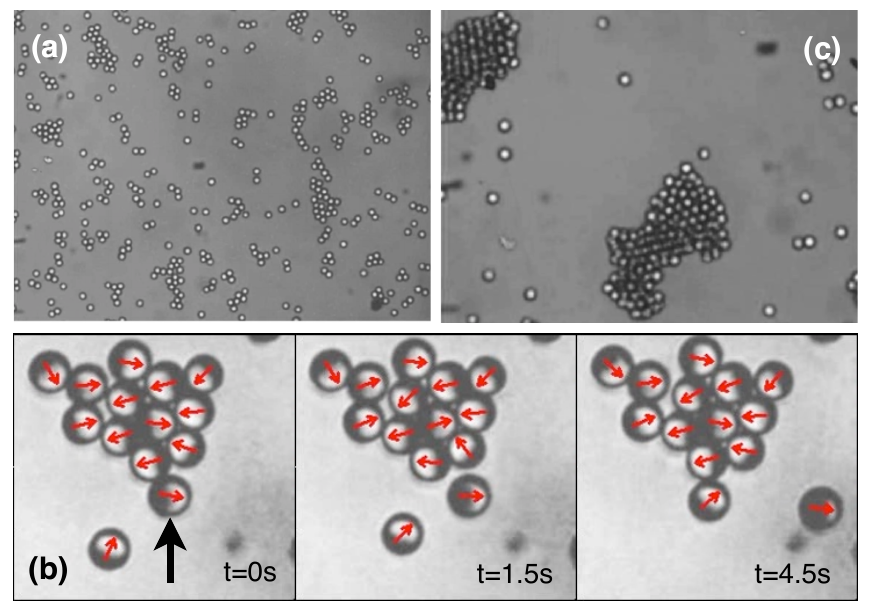}
  \caption{Carbon-coated colloidal self-propelled particles in a
    locally demixing  water-lutidine mixture: (a)~Cluster formation at low
    densities ($\phi\simeq0.1$). (b)~Resolved particle orientations and
    observation of particle interchange. If the particles rotational diffusion
    is fast enough, it escapes an initial cluster before other particles join
    the
    cluster. The snapshots show one such event, where a particle (arrow)
    leaves the cluster and is replaced by another particle. (c)~Evidence for
    phase separation at higher densities. Figure adapted from
    Ref.~\citenum{butt13}.}
  \label{fig:self_trapping}
\end{figure}

Buttinoni~\textit{et al.} have used a different experimental setup of
colloidal self-propelled Janus particles that are, however, not driven by
chemical reactions~\cite{volpe11,butt12,butt13}. The spherical particles are
prepared from silica beads, where one hemisphere is coated with carbon. Here,
particles are confined between two glass slides in a quasi two-dimensional
geometry and are suspended in a water-2,6-lutidine mixture, which at room
temperature is just below the critical temperature $\sim33^\circ$C. When the
suspension is illuminated by a widened laser beam (532 nm), the carbon absorbs
the light and the solvent is locally heated above the critical
point. Consequently, the solvent demixes locally at the carbon side of the
particles. The particles behave as Brownian self-propelled particles in two
dimensions with a propelling speed that is proportional to the light
intensity~\cite{volpe11,butt12}. The propulsion mechanism is
diffusiophoresis~\cite{butt12}. Carbon has been employed as light-absorbing
material since its Hamaker constant is substantially lower compared to gold
(or any other metal). Attractive forces in the passive suspension are thus
almost negligible as demonstrated by the measured pair distribution
function~\cite{butt13}. Another advantage of this setup compared to catalytic
colloidal swimmers is that for the considered light intensities phoretic
attraction can also be neglected. While it has been shown that, in principle,
there exists a phoretic attraction between particles for sufficient high light
intensities, active suspensions have been studied at intensities far below
this threshold. This has been tested by measuring the pair distribution
function for spherical passive particles in the vicinity of a Janus particle
stuck to the glass slide both in and out of equilibrium, whereby no
qualitative deviations have been observed.

Again, the experiment of Buttinoni~\textit{et al.} shows the formation of
clusters as soon as the particles are activated (see
Fig.~\ref{fig:self_trapping}). At low densities, a linear
relation between mean cluster size and particle speed $v_0$ is found similar
to the other two experiments in Refs.~\citenum{theur12,pala13}. The clustering
mechanism can be described as the
competition between two time scales, which has also been done in
Refs.~\citenum{redner13_prl,redner13_pre} in terms of
a kinetic model. The physical picture is that of colliding particles, which
block each other (``self-trapped'') due to the persistence of their motion,
where orientations decorrelate on time scales $\sim1/\Dr$ with rotational
diffusion coefficient $\Dr$. If this rotational diffusion is slow enough
compared to the mean free time, other particles may join the cluster before the
initial particles are able to escape the ``seed'', cf.
Fig.~\ref{fig:self_trapping}(b). After sufficient time, a dynamic steady state
should be reached, where on average just as many particles escape the cluster as
new particles join the cluster. Clusters are indeed very dynamic objects, where
particles are interchanged continuously, see Fig.~\ref{fig:self_trapping}.
Moreover, it has been possible to resolve particle orientations so that the
self-trapping mechanism could be confirmed qualitatively,
cf. Fig.~\ref{fig:self_trapping}(b). When illumination is turned off, clusters
dissolve until the system reaches thermal equilibrium.

At higher densities, the experiment of Buttinoni~\textit{et al.} shows a gas
phase with a few big and
slowly moving clusters, see Fig.~\ref{fig:self_trapping}(c). Following the
passive scenario of phase separation, one might expect the final state to
contain one single large cluster. However, larger clusters move very slowly so
that the actual merging of all clusters is not observed within the
experimental time window. For example, clusters in the experiment are not
perfectly two dimensional objects (they might ``buckle'' out-of-plane) and
approaching the cell walls they slow down. Nevertheless, while monitoring a
region consisting $N$ particles, all particles in clusters larger than $N/10$
have been added up and the fraction of total particles in a cluster is
identified as the order parameter $P$. By measuring this order parameter for
different propelling speeds and densities, one observes a continuous increase,
which is moreover supported by Brownian dynamic simulations as discussed in
the next section. The transition occurs at lower densities than predicted by
the simulations of perfectly hard disks. Still, the critical swimming speeds
obtained from experiments and simulations at intermediate packing fraction
$\phi=0.36$ coincide quite well.

Although each experiment shows an individual method to prepare self-propelled
particles, we observe dynamical clustering to be quite generic. To gain
further insight into the phase behavior, we seek assistance from analytical
and numerical work as detailed in the next section.


\section{Model and numerical results}

\subsection{Model}
\label{sec:model}

In order to analytically and numerically study suspensions of self-propelled
colloidal particles, one needs a suitable minimal model for the
experiments discussed in the previous section that is both simple and
tractable, but contains the relevant physics. Assuming that the dynamics is
overdamped as appropriate for solvated colloidal particles at low Reynolds
numbers, the Langevin equation is applicable, i.e.,
\begin{equation}
  \label{eq:lang}
  \dot\x_i = -\mu_0\nabla U + v_0\vec e_i + \nois_i.
\end{equation}
The mobility of a free particle is denoted by $\mu_0$. The noise term
$\nois_i$ models the thermal motion and has zero mean and variance
\begin{equation} 
\langle\nois_i(t)\nois_j^T(t')\rangle=2D_0\mu_0^{-2}\delta_{ij}
\delta(t-t') ,
\end{equation}
with $D_0=k_BT\mu_0$ denoting the bare diffusion coefficient and $k_BT$ the
thermal energy. Particles are restricted to two dimensions and interact via a
pair potential $u(r)$, where the total energy is given by $U=\sum_{i<j}u(|\vec
r_i - \vec r _j|)$. Each particle has an orientation $\vec e_i = (\cos
\varphi_i , \sin \varphi_i)$, along which the particle is propelled with
constant speed $v_0$. Of course, the model does not resolve the microscopic
origin of the directed motion but requires $v_0$ as an input parameter. The
orientational angle $\varphi_i$ fluctuates freely with diffusion coefficient
$\Dr$ according to
\begin{equation}
  \langle\dot\varphi_i\rangle =0, \quad
  \langle\dot\varphi_i(t)\dot\varphi_j(t')\rangle = 2\Dr\delta_{ij}
  \delta(t-t').
\end{equation}
On time scales $\gg1/\Dr$, the motion of a \emph{single} propelled particle
becomes effectively diffusive with increased long-time diffusion coefficient
$D_\text{eff}=D_0+v_0^2/(2\Dr)$~\cite{howse07}, making it possible to define an
effective temperature $\Te\sim v_0^2$, which is strongly
modified for interacting particles~\cite{bialke12,fily14}. In case of free
particles it has been shown that particles being trapped in a harmonic external
potential, do not follow the concept of an effective temperature, while the
sedimentation of free self-propelled particles is describable in terms of an
effective temperature~\cite{szamel14,tailleur09,pala10}. In the following, we
now review the key results from numerical studies of the particle model based on
Eq.~\eqref{eq:lang}.

\subsection{Freezing}

The model just described has been used first by Bialk\'e~\textit{et al.} to
study the freezing transition of an active suspension at high
densities~\cite{bialke12}.
Particles interact via the Yukawa pair potential $u(r)=\Gamma e^{-\lambda r}/r$
with fixed inverse screening length $\lambda=3.5$ leaving the coupling strength
$\Gamma$ and the free swimming velocity $v_0$ as free parameters. By applying
both static and dynamic criteria for the freezing and melting, it is shown that
the suspension is first ordered structurally before dynamical freezing can be
observed. As structural measure the local hexagonal bond-orientational order has
been evaluated, which is quantified through
\begin{equation}
  \label{eq:bo}
  q_6(i)= \frac{1}{|\mathcal{N}(i)|}\sum\limits_{j\in
    \mathcal{N}(i)}e^{i6\theta_{ij}}.
\end{equation}
Here, $\theta_{ij}$ is the angle enclosed between the displacement vector of
particles $i$ and $j$ and a fixed axis, and $\mathcal{N}(i)$ is the set of the
neighbors of particle $i$, usually within a threshold distance. By averaging
$q_6$ over all particles and squaring its absolute value, one gets a global
structural order parameter which is $0$ for an unordered suspension and $1$ for
a perfect hexagonal crystal. Note that although a large cluster might show high
crystalline order, this global parameter is still $0$ due to the particles in
in the gas phase and the crystalline domains within the cluster which are tilted
to one another and separated by linear defects~\cite{redner13_prl}. However, by
increasing the propulsion speed, the transition to a hexagonal crystal is
shifted towards higher critical coupling strengths $\Gamma_c\sim\sqrt{\phi}/T$.
Another result of this work has been that, similar to the clustering transition,
the shifted freezing transition cannot be described by an effective temperature
$\Te\sim v_0^2$. In related studies, Berthier \textit{et
  al.}~\cite{berthier_arxiv,berthier13} have shown that a glass forming system
exhibits a shift towards higher temperatures for the kinetic arrest to occur
if particles are active. It appears that this shift cannot be explained by the
simple picture of an effective temperature, but allows the study of glasses at
high packing fractions~\cite{ni13}. In yet another numerical study for a soft
interaction potential of a polydisperse suspension, Fily \textit{et
  al.}~\cite{fily14} have resolved the complete phase diagram, where a fluid,
a phase separated regime and, due to the polydispersity, a glassy regime is
identified, see Fig.~\ref{fig:phase}(b).

\subsection{Clustering}
\label{sec:clustering}
\begin{figure}
  \includegraphics[width=\linewidth]{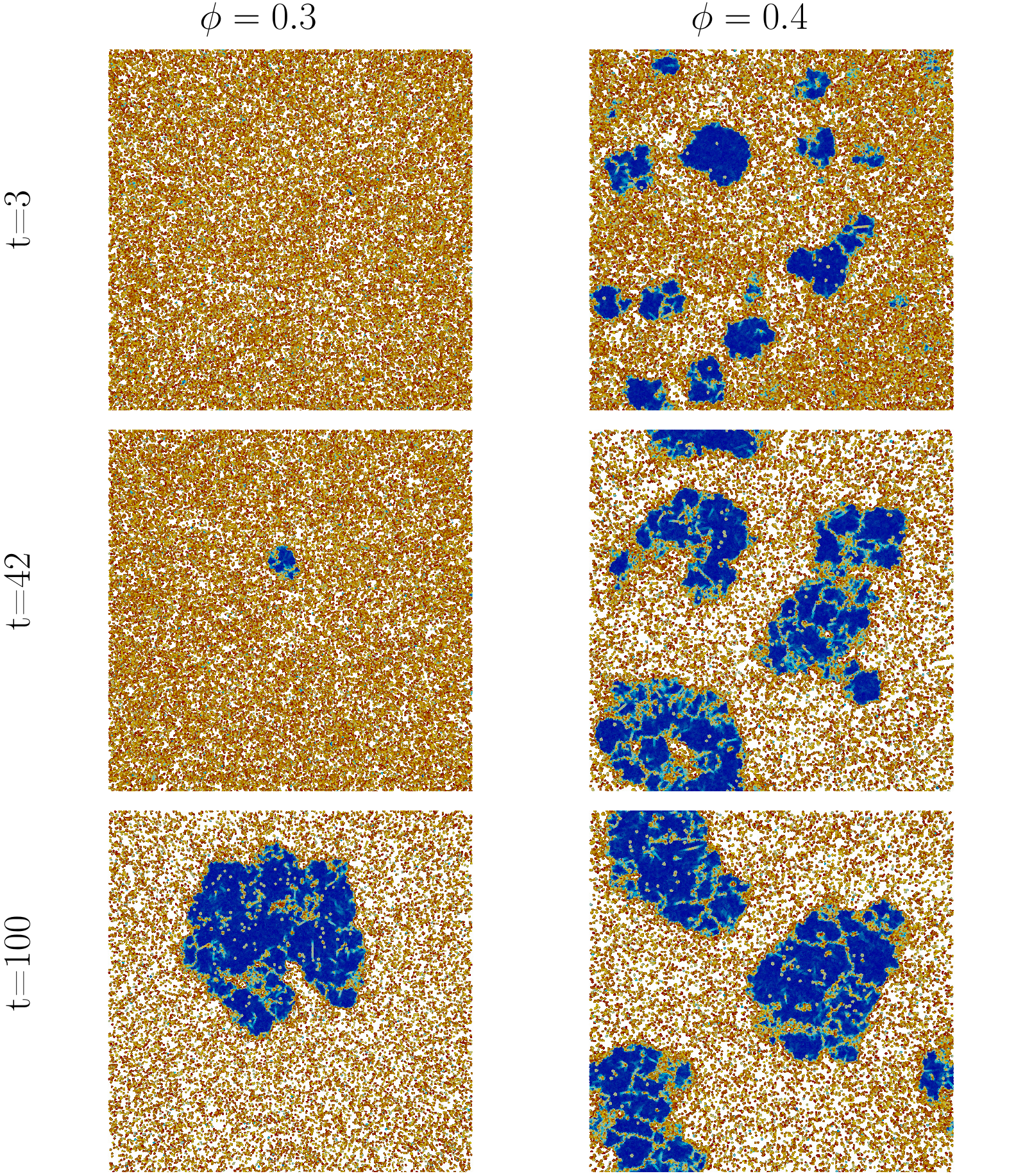}
  \caption{Phase separation dynamics for the minimal model at two densities:
    area fraction $\phi=0.3$ (left) and $\phi=0.4$ (right). The snapshots show
    particle-resolved simulations for $N=40000$ particles at three different
    times given in units of a typical Brownian time. The suspension is
    equilibrated at $v_0=0$ and then quenched instantaneously to $v_0=100$.
    At lower density we observe a nucleation-type scenario: after a delay one
    single cluster starts to grow until the steady state is reached. In
    contrast, at a higher density multiple domains form immediately after the
    quench and then grow and merge until eventually a single dense droplet is
    reached. This scenario is usually described as spinodal decomposition.
    Particles are encolored according to their $q_6$ value, where red
    particles correspond to $q_6=0$ and blue particles to $q_6=1$.}
  \label{fig:snap}
\end{figure}

By now, extensive numerical simulations of the minimal model have been
performed by several groups employing different repulsive pair
potentials~\cite{fily12,sten13,redner13_prl,butt13,bialke13,sten14,fily14,
speck14}. The clustering and phase separation of athermal self-propelled
particles has been reported first by Fily~\textit{et al.} in
Ref.~\citenum{fily12} employing the minimal model. For particle interactions,
the authors have chosen a non-diverging pair potential, i.e., harmonic repulsion
in case of particle overlap.  Moreover, the authors neglect translational noise
($D_0=0$) and treat rotational diffusion as an independent fixed parameter. They
perform molecular dynamics simulations of a monodisperse suspension with up to
10000 particles.
They show that systems above $\phi\approx0.4$ phase separate into one big
cluster surrounded by a gas phase. The experimentally observed clustering at
lower densities (Refs.~\citenum{theur12,pala13}) and phase separation into a few
big and slow clusters (Ref.~\citenum{butt13}) are not
observed in the simulations. However, in qualitative agreement with the
experiment by Palacci \textit{et al.}~\cite{pala13}, giant number fluctuations
have been reported for suspensions above the critical density. Furthermore, the
behavior cannot be mapped to a system with an effective temperature $\Te\sim
v_0^2$ in agreement with~\cite{bialke12}. Finally, in accordance with the
experiment by Theurkauff~\emph{et al.}, an apparently diverging behavior of the
static structure factor for $k\rightarrow0$ is found for phase separated
systems.

Even larger systems with up to 512000 particles have been simulated by
Redner~\textit{et al.}~\cite{redner13_prl}. In contrast to the previous work,
particles interact through the WCA potential~\cite{weeks71} as appropriate for
hard colloidal particles. Translational noise is included and the rotational
diffusion coefficient is coupled to translational diffusion via the
Stokes-Einstein-Debye relation $\Dr=3D_0/a^2$, where $a$ is the particle
diameter which is defined through the potential. By varying the packing fraction
$\phi$ and the propulsion speed $v_0$, extensive simulations have lead to a
phase diagram characterized by the fraction of particles in the dense cluster
phase. Again, similar to Fily~\textit{et al.} in Ref.~\citenum{fily12}, a
clustering transition is observed. Moreover, a remarkable result is that a
simple model of rate equations for particles joining and leaving a given cluster
shows excellent agreement with the numerical data. The authors also studied
structural properties within the dense phase through bond-orientational order,
cf. Eq.~\eqref{eq:bo}, where one notices 5-fold and 7-fold point defects as well
as linear defects separating crystalline domains within the cluster. The authors
also measure the spatial correlation of the bond-orientational order parameter
in large clusters, where they observe a transition from liquid-like exponential
decay to a hexatic-like power-law decay as they increase the swimming speed,
which is similar to the freezing by heating transition observed by Helbing
\textit{et al.} \cite{helbing00}.
Furthermore, different phase separation scenarios are observed: on the one
hand the system shows delayed nucleation like an equilibrium system near the
binodal and one observes the growth of one single cluster. On the other hand,
a denser system shows spinodal-like coarsening behavior with several clusters
(see Fig.~\ref{fig:snap} for data obtained for a similar system). Redner
\textit{et al.} also report that the asymptotic growth of the mean cluster size
follows $\sim t^{1/2}$. However, the value of the exponent has to be treated
with care and more recent results indicate an asymptotic value $1/3$, which is
also expected for passive phase-separated
suspensions~\cite{wittkowski14,speck14}.

\begin{figure}[t]
  \includegraphics{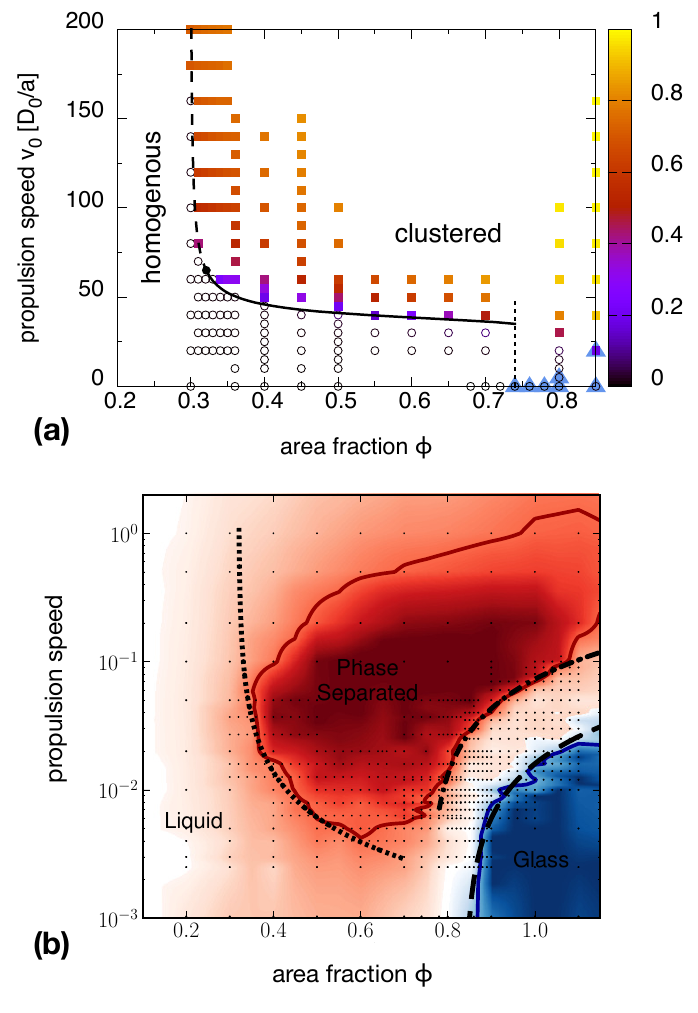}
  \caption{~Numerical phase diagrams: (a)~For a monodisperse active suspension
    in which particles interact via the short-ranged repulsive WCA
    potential. The color scale corresponds to the
fraction    of particles in the dense phase, whereby open symbols indicate a
    homogeneous suspension and
    closed symbols indicate the clustered phase. Also shown is an prediction
    for the instability line (solid and dashed line). The vertical dashed line
    indicates the equilibrium freezing density with triangles corresponding to
    the solid state as identified from the bond-orientational order parameter
    Eq.~\eqref{eq:bo}. (b)~For a polydisperse active suspension (neglecting
    translational noise) employing a soft repulsive pair potential. Here a
    glassy region instead of freezing is observed. Red regions correspond to
    systems with high number fluctuations, while blue ones show systems with
    slow dynamics. Adapted from Ref.~\citenum{speck14} and arXiv:1309.3714v1.}
  \label{fig:phase}
\end{figure}

In Fig.~\ref{fig:phase} two numerical phase diagrams for the minimal model are
presented. In Fig.~\ref{fig:phase}(a) results employing the hard WCA potential
are shown, where a range of state points $(\phi,v_0)$ have been simulated. In
Fig.~\ref{fig:phase}(b), a similar model has been studied but without
translational noise and employing a much softer repulsive potential, that
allows particles to overlap. Moreover, the particle sizes are not identical
but drawn from a distribution. As mentioned, this leads to a qualitative
change at high densities with the appearance of a glassy phase. Together,
these results demonstrate that the described phase separation in active
suspensions with a purely repulsive pair potential is a robust phenomenon that
does not depend so much on the interaction details.

\subsection{Hydrodynamic effects}

Although the numerical results shown have been obtained using the simple
particle model given by Eq.~\eqref{eq:lang}, one already observes
qualitatively quite good agreement with the experiments. For a more faithful
modeling of the experimental setups one needs to include hydrodynamic
interactions, not only between particles, but also between particles and
confining walls. One promising direction is the hydrodynamic model derived by
Ishikawa \textit{et al.}, which prescribes the tangential surface velocity
$\vec v_i^s$ of the fluid at swimmer $i$ according to
\begin{equation}
  \vec v_i^s=B_1(1+\beta\vec e_i\cdot\vec r_i^s)[(\vec e_i\cdot\vec
  r_i^s)\vec r_i^s-\vec e_i],
\end{equation}
where $\vec r_i^s$ denotes the normalized vector pointing from the particle
center to a surface point~\cite{ishikawa06}. This type of hydrodynamic
swimmers are called \textit{squirmers}. The free swimming velocity is
proportional to $B_1$, while the factor of proportionality depends on the
spatial dimensions of the system. The quantity $\beta$ determines the symmetry
of the velocity field and characterizes a particle with $\beta<0$ as
``pusher'' and with $\beta>0$ as ``puller''~\cite{blake71}.
For $\beta=0$, the velocity field at the particle surface is symmetric and the
particle can be considered as a neutral squirmer. For the connection between
propulsion mechanism and the squirmer model, see, e.g., Ref.~\citenum{yang14}.

For suspensions of active particles, the required computational power limits
the total number of particles that can be simulated to currently a few
hundreds so that results have to be analyzed carefully regarding finite size
effects. Ishikawa and Pedley~\cite{ishikawa08} have simulated up to 196
squirmers restricted to two-dimensional motion in a monolayer within in an
unbounded three-dimensional fluid. Although particles tend to align, which
counters the self-trapping mechanism discussed before, they observe the
formation of clusters. In addition they have considered bottom-heavy
particles, i.e., particles with a shifted center of mass, which tend to swim
upwards and are able to prevent sedimentation~\cite{wolff13}. In this case the
formation of bands is observable. Another work by
Fielding~\cite{arxiv_fielding} considers 256 neutral squirmers restriced to
two dimensions in a two-dimensional fluid. It is shown that phase separation
is strongly suppressed due to hydrodynamic interactions. The mechanism
responsible for the suppression is an effective hydrodynamic torque turning
particle orientations so that head-on collisions (and thus the trapping time)
are reduced. More recently, Z\"ottl and Stark~\cite{zott14} have considered
208 squirmers moving in strong confinement. In case of $\beta \neq 0$ they
also found that phase separation is suppressed. However, neutral squirmers
phase separate more clearly into a crystal phase and a gas phase than
particles modelled by Eq.~\eqref{eq:lang}. This effect is caused by a slow
down due to hydrodynamic interaction between particles and hydrodynamic
swimmer-wall interactions. The authors show that the angular distribution of
the squirmers is broadened and particles also tend to orient perpendicular to
the cell wall thus enhancing the self-trapping mechanism. This could be one of
the reasons why experimental systems tend to cluster at lower densities than
observed in the Brownian dynamics simulations neglecting hydrodynamics.


\section{Mean-field theory}
\label{chap:mean_field}

\subsection{Derivation}

We now briefly sketch the systematical derivation of the coupled mean-field,
effective hydrodynamics equations of motion developed in
Ref.~\citenum{bialke13}. As starting point, we note that an equivalent
description of the numerical model given by Eq.~\eqref{eq:lang} is provided
through the Smoluchowski equation
\begin{equation}
\label{eq:smoluchowswki}
 \partial_t\Psi_N = \sum\limits_{i=1}^N\nabla_i\cdot[(\nabla_i U)-v_0\vec e_i +
\nabla_i]\Psi_N + \Dr \sum\limits_{i=1}^N
\frac{\partial^2\Psi_N}{\partial\vhi^2_i},
\end{equation}
where $\Psi_N(\{\vec r_i ,\vhi_i\},t)$ is the joint probability distribution of
all possible configurations. Since particles are identical, one random
particle is tagged and the subscript for position and orientation is dropped.
Then, $\Psi_N$ is integrated over all other particle positions and orientations
to obtain the one particle probability distribution $\Psi_1(\vec r,\vhi,t)$.
Its evolution obeys
\begin{equation}
\label{eq:psi1}
 \partial_t \Psi_1=-\nabla\cdot[\vec F + v_0\vec e\Psi_1 -\nabla\Psi_1] + \Dr
\partial^2_\vhi\Psi_1,
\end{equation}
where $\vec F$ is the mean force acting on the tagged particle, which depends
on higher many-body probability distributions and leads to the well-known
BBGKY hierarchy~\cite{hansen06}.

As a closure already on the level of the single particle density, we project
the mean force onto the orientation of the tagged particle, $\vec
F\approx(\vec e\cdot\vec F)\vec e$ and introduce an effective diffusion
coefficient (for details see Ref.~\citenum{bialke13}). We thus find the
mean-field evolution equation for the one particle density
\begin{equation}
  \label{eq:1}
  \partial_t\Psi_1 = -\nabla[v(\rho)\vec e - D\nabla]\Psi_1 +
  \Dr\partial_\vhi^2\Psi_1.
\end{equation}
Here, $\rho$ denotes the local density, $D$ the long time diffusion
coefficient in a passive suspensions, and
\begin{equation}
  \label{eq:v}
  v(\rho) = v_0 - \rho\zeta
\end{equation}
is the effective swimming speed. Here, we have assumed that the local density
$\rho(\vec r,t)$ is a sufficiently slowly varying field so that we can replace
the homogeneous density $\bar{\rho}$ with the local density $\rho(\x,t)$. This
assumption holds close to the onset and during the initial stages of the
dynamical instability. Although a linearly decaying $v(\rho)$ has been
considered before~\cite{thomp11}, Bialk\'e~\textit{et al.} have shown the
derivation from first principles. 

\begin{figure}[t]
  \includegraphics{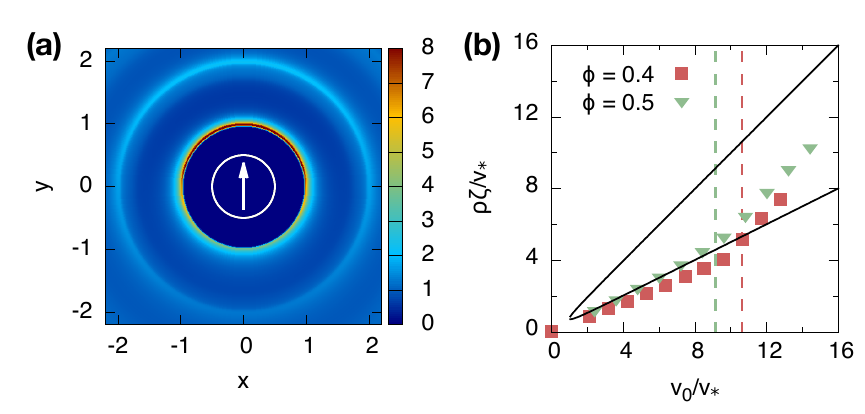}
  \caption{(a)~Anisotropic pair distribution function from numerical
    simulations using the WCA potential plotted in the $xy$-plane. The white
    circle represents the tagged particle with the white arrow indicating the
    particle orientation. (b)~Plot of the normalized force imbalance
    coefficient $\bar\rho\zeta/v_\ast$ as a function of the reduced swimming
    speed $v_0/v_\ast$ for two area fractions $\phi$. The characteristic speed
    is set by $v_\ast=4\sqrt{D\Dr}$. The dashed vertical lines correspond to
    the phase transition points determined with the help of finite-size
    scaling. The two solid lines represent the boundaries of the instability
    region determined by Eq.~\eqref{eq:unstable_zeta}. Adapted from
    Ref.~\citenum{bialke13}.}
  \label{fig:anisotropic}
\end{figure}

The effective swimming speed given by Eq.~\eqref{eq:v} is reduced due to the
interactions with other particles as quantified by the force imbalance
coefficient
\begin{equation}
  \label{eq:zeta}
  \zeta = \IInt{r}{0}{\infty} r[-u'(r)] \IInt{\theta}{0}{2\pi}
  \cos\theta\,g(r,\theta),
\end{equation}
where the prime denotes the derivative with respect to the argument and $\theta$
is the angle enclosed between the orientation of the tagged particle and the
displacement vector from the tagged particle to another particle at distance
$r$. The physical picture behind Eq.~\eqref{eq:v} is that particle collisions
are more likely occuring in the direction of propulsion, causing an anisotropic
two-dimensional pair distribution function $g(r,\theta)$. This picture is
confirmed in computer simulations, see Fig.~\ref{fig:anisotropic}(a). While we
found a linear relationship Eq.~\eqref{eq:v}, Cates and coworkers have
considered more general functional
dependences $v(\rho)$ and have shown that a reduced particle mobility in dense
regions might lead to further accumulation of particles in these regions and
finally to phase separation~\cite{tailleur08,cates10,cates13}. Note that the
competition of time scales is also reflected in
Eq.~\eqref{eq:v}. The mean collision rate is connected to the density $\rho$
and swimming speed $v_0$, while the rate of reorientation influences the value
of $\zeta$. This is demonstrated in the limit $\Dr\rightarrow\infty$, where
orientational fluctuations are so fast that on average the pair distribution
function is isotropic and the force imbalance vanishes, $\zeta=0$.

The local density $\rho(\vec r,t)$ and the orientational field $\vec p(\vec
r,t)$ are given by
\begin{equation}
  \rho(\x,t) = \IInt{\vhi}{0}{2\pi} \psi_1(\x,\vhi,t)
\end{equation}
and the first moment
\begin{equation}
  \vec p(\vec r,t) = \IInt{\vhi}{0}{2\pi} \vec e\,\psi_1(\x,\vhi,t),
\end{equation}
respectively. The equations of motion for these two fields become the
continuity equation
\begin{equation}
  \label{eq:meanfieldrho}
  \partial_t\rho = -\nabla\cdot[v\vec p -D\nabla\rho]
\end{equation}
for the density and, through neglecting the coupling to second harmonics of
the orientational angle,
\begin{equation}
  \label{eq:meanfieldp}
  \partial_t\vec p = -\frac{1}{2}\nabla(v\rho) + D\nabla^2\vec p - \Dr \vec p.  
\end{equation}
The first term on the right hand side can be interpreted as an effective
pressure $P(\rho)=\frac{1}{2}v(\rho)\rho$, the second term is akin to a
viscosity term, and the last term describes the local relaxation due to the
rotational diffusion. While these equations have been derived systematically
from the Smoluchowski equation~\eqref{eq:smoluchowswki}, they can also be
obtained from the phenomenological equations of Toner and Tu~\cite{tone95}
(see, e.g., Ref.~\citenum{fily14}) by neglecting all higher order terms. Of
course, in this case the various coefficients are in principle unknown.

\subsection{Dynamical instability}

The linear stability of Eqs.~\eqref{eq:meanfieldrho} and \eqref{eq:meanfieldp}
against density fluctuations has been studied by Speck~\textit{et al.} and
Fily~\textit{et al.}~\cite{fily12,fily14,speck14,bialke13}. By considering large
length scales and time
scales much longer than $1/\Dr$, Eq.~\eqref{eq:meanfieldp} can be written as
\begin{equation}
  \label{eq:p_adiabtic}
  \vec p \approx -\frac{1}{2\Dr}\nabla(v\rho),
\end{equation}
so that the orientational field is adiabatically connected to the density
field. By putting this expression into Eq.~\eqref{eq:meanfieldrho}, we obtain
the diffusion equation
\begin{equation}
  \partial_t \rho = \nabla \mathcal{D}(\rho) \nabla \rho.
\end{equation}
While we have thus eliminated the orientational field $\vec p$, the effects of
the force imbalance are retained through the effective swimming speed and give
rise to a density-dependent, collective diffusion coefficient
\begin{equation}
  \mathcal{D}(\rho)=D+\frac{(v_0-\rho\zeta)(v_0-2\rho\zeta)}{2\Dr}.
\end{equation}
If $\mathcal{D}(\rho)<0$ the system becomes locally unstable and density
fluctuations grow exponentially until they saturate due to the coupling to
nonlinear modes. The criterion $\mathcal{D}(\rho) < 0$ is fulfilled if
$\zeta_-\leqslant\zeta\leqslant\zeta_+$ with
\begin{equation}
  \label{eq:unstable_zeta}
  \frac{\bar{\rho}\zeta_\pm}{v_\ast}=\frac{3}{4}(v_0/v_\ast)\pm\frac{1}{4}
  \sqrt{(v_0/v_\ast)^2 - 1}.
\end{equation}
This result implies that at least a propulsion speed $v_0>v_\ast=4\sqrt{D\Dr}$
is necessary for the instability to occur, see Fig.~\ref{fig:anisotropic}(b).

This prediction has been tested for numerical results employing the WCA
potential, see Fig.~\ref{fig:anisotropic}(b). For two densities, the
transition speeds $\vc$ have been estimated with the help of finite-size
scaling of an order parameter, in our case the mean fraction of particles in
the largest cluster~\cite{bialke13}. For each propulsion speed $v_0$ we have
also determined the force imbalance through Eq.~\eqref{eq:zeta} from the
measured pair distribution function, and the passive long-time diffusion
coefficient $D$ at that density, which determines $v_\ast$. The result in
Fig.~\ref{fig:anisotropic}(b) shows good agreement between the estimated
transition speeds $\vc$ and the crossing of the imbalance coefficient into the
instable region.


\section{Concluding remarks}

To conclude, we have reviewed experimental, numerical, and analytical work on
the phase behavior of self-propelled colloidal particles without explicit
alignment interactions in (quasi) two dimensions. Although different physical
mechanisms are responsible for the ``swimming'' of particles in the three
experimental setups, clustering of particles is observed to be a generic,
robust feature of active suspensions. Supported by computer simulations of a
minimal model, it has been established that the self-propulsion of repulsive
particles is able to induce a phase separation into dense and dilute
regions. In the mean-field picture, this phase separation can be explained as
a dynamical instability, where in the dilute regions the fast particles exert
an effectively higher pressure compared to dense but slow regions. Although
active suspensions are genuinely out-of-equilibrium systems, this phenomenon
is surprisingly similar to liquid-vapor phase separation in a passive
suspension with sufficiently strong attractive interactions. At higher
densities, the self-propulsion shifts the freezing (or glass-forming)
transition. In particular, at fixed density the suspension is melted before
entering the phase-separated state as the propulsion speed is increased.

The pivotal role in the mean-field theory is played by the force
imbalance. This takes into account that the pair distribution function is not
isotropic with respect to particle orientations although particles are
spherical. In computer simulations, we found good agreement with the
mean-field prediction for the onset of the dynamical instability.

Departing from the minimal model, there are several direction into which
further studies might go. Recently, first studies have started to investigate
the phase behavior of the minimal model in three
dimensions~\cite{sten14,wyso14} and the influence of attractive
forces~\cite{redner13_pre,mognetti13} or non-spherical particle
shape~\cite{wensink12_jpcm,wensink08}. Active particles (bacteria or colloidal
particles) with entropic attraction due to depletants (polymers) have been
studied in Ref.~\citenum{line12} and Ref.~\citenum{das14}, showing that phase
separation is suppressed if alignment interactions are introduced. For
catalytic swimmers, the effects of local fuel depletion leading to collective
chemotactic behavior have started to receive
attention~\cite{meyer14,pohl14,soto14}.

Another open issue is the more faithful modeling of the experiments in order
to achieve more than qualitative agreement. An important step is to include
hydrodynamic interactions. While the squirmer model seems to be a good
starting point, it is not yet clear to which extent it reproduces the actual
flow pattern of self-propelled colloidal particles in particular in dense
suspensions. Moreover, accessible system sizes are still rather small. From
the experimental side, it would be highly desirable to clarify to what degree
the propulsion of a particle within a (quasi-two-dimensional) cluster is
comparable to free propulsion. This might seem questionable for both the
chemically driven particles consuming H$_2$O$_2$ as well as the particles
driven by reversible demixing of the solvent. In the first case H$_2$O$_2$
might be depleted while for the second case shared demixing zones within
clusters develop, leading to a reduction of directed motion.

But even for the minimal model Eq.~\eqref{eq:lang} there is plenty of work
left to do. For phase transitions in passive systems an elaborate framework
has been developed over the years, which allows to reliably construct phase
diagrams and to study critical phenomena in finite-size computer
simulations. Not much is known yet for active suspensions.

\acknowledgments

We gratefully acknowledge support by the Deutsche Forschungsgemeinschaft
through the recently established priority program 1726.


\bibliography{refs}

\end{document}